# Clustering Time Series Data Stream – A Literature Survey

V.Kavitha ,.M.Punithavalli

[†] *Computer Science Department, Sri Ramakrishna College of Arts and Science for Women,Coimbatore,Tamilnadu,India.*
[††] *Sri Ramakrishna College of Arts & Science for Women, Coimbatore ,Tamil Nadu, India.*

Kavithaanand11@gmail.com ,mpunitha_srcw@yahoo.co.in

***Abstract***-Mining Time Series data has a tremendous growth of interest in today's world. To provide an indication various implementations are studied and summarized to identify the different problems in existing applications. Clustering time series is a trouble that has applications in an extensive assortment of fields and has recently attracted a large amount of research. Time series data are frequently large and may contain outliers. In addition, time series are a special type of data set where elements have a temporal ordering. Therefore clustering of such data stream is an important issue in the data mining process. Numerous techniques and clustering algorithms have been proposed earlier to assist clustering of time series data streams. The clustering algorithms and its effectiveness on various applications are compared to develop a new method to solve the existing problem. This paper presents a survey on various clustering algorithms available for time series datasets. Moreover, the distinctiveness and restriction of previous research are discussed and several achievable topics for future study are recognized. Furthermore the areas that utilize time series clustering are also summarized.

## I.INTRODUCTION

Today Time Series data management has become an interesting research topic by the data miners. Particularly, the clustering of time series has attracted the interest of researchers. Data mining is usually constrained by three limited resources. They are Time, Memory and Sample size. Recently time and memory seem to be bottleneck for machine learning application. Clustering is an unsupervised learning process for grouping a dataset into subgroups. A data stream is an ordered sequence of points $x_1, , , , , ,x_n$. These data can be read or accessed only once or a small number of times. A time series is a sequence of real numbers, each number indicating a value at a time point. Data flows continuously from a data stream at high speed, producing more examples over time in recent real world applications. Traditional algorithms cannot support to the high speed arrival of time series data. This is a reason; the new algorithms have been developed for real time processing data.

Time series data are being generated at an unique speed from almost every application domain e.g., Daily fluctuations of stock market, Fault diagnosis, Dynamic scientific experiments, Electrical power demand, position updates of moving objects in location based services, various reading from sensor networks, Biological and Medical experimental observations, etc. Traditionally clustering is taken as a batch procedure. Most of the clustering techniques can be two major categories. One is Partitional clustering and another one is Hierarchical Clustering [1]. They are the two key aspects for achieving effectiveness and efficiency when using time series data. A time series experiment requires multiple arrays which all makes it very expensive. Dimensionality reduction techniques can be divided into two groups (i) Feature Extraction (ii) Feature Selection. Feature Extraction techniques extract a set of new features from the original attributes. Feature Selection is a process that selects a subset of original attributes. There have been numerous textbooks [5] and publications on clustering of scientific data for a variety of areas such as taxonomy, agriculture [2], remote sensing [3], as well as process control [4]. This paper presents a survey on various clustering algorithms available for time series datasets. Moreover, the distinctiveness and restriction of previous research are discussed and several achievable topics for future study are recognized. Furthermore the areas that time series clustering have been applied to are also summarized.

The remainder of the paper is organized as follows. Section 2 reviews the concept of time series and gives an overview of the algorithms of different techniques. Section 3 marginally discusses possible future extensions of the work. Section 4 concludes the paper with fewer discussions.

## II.RELATED WORK

Quite a number of clustering techniques has been proposed earlier for time series data streams. This section of the paper discusses some of the earlier proposed methods for efficient clustering of time series datasets.

Ville Haulamati et al. in [8] poses problem related to time series data clustering in Euclidean space using Random Swap (RS) and Agglomerative Hierarchical clustering followed by k-mean fine-tuning algorithm to compute locally optimal prototype. It provides best clustering accuracy. And also provide more improvement to k-medoids. The drawback of this algorithm is, it outperforms the quality.

Pedro Pereiva Rodrigous et al. in [6] analyzes an incremental system for clustering streaming time series, using Online Divisive Agglomerative Clustering system continuously maintains a tree-like hierarchy of clusters using a top-down strategy. Using ODAC cluster quality is to be measure to calculate cluster's diameter. The highest dissimilarity between objects of the same cluster is defined as diameter. The strength of ODAC is do not need a predefined number of target clusters. It provides a good performance on finding the correct number of clusters obtained by a bunch of runs of k-Means. The disadvantage of this system is when the tree structure expands, the variables should move from root to leaf, when there is no statistical confidence on the decision of assignment may split variables. And the computation of high dimensional data being processed may represent a drawback of the clustering procedure.

Xiang Lian et al. in [7] proposed that all types of time series data applications needs an efficient and effective similarity search over stream data is essential. To predict the unknown values that have not arrived at the system and answer similarity queries based on the predicted data using the three approaches namely Polynomial, discrete Fourier Transform (DFT) and Probabilistic. These approaches can lead to good offline prediction accuracy and not suitable for online stream environment. Because online requires low prediction and training costs. These approaches are straight forward for seeking general solutions. And it gives proper confidence for prediction. It can predict values while explicitly providing a confidence. The polynomial approach that predicts future values based on the approximated curve of recent values. The Discrete Fourier Transform (DFT) forecasts the future values using approximations in the frequency domain. And the probabilistic approach can provide predicting values and it can be adaptive to the change of data. The group probabilistic approach is utilizing the correlations among stream time series. The drawback of this probabilistic approach, it needs more time to predict the future values.

Sudipto Guha et al. in [13] described a streaming algorithm that effectively clusters large data streams. For analysis of such data, the ability to process the data in a single pass, or a small number of passes, while using little memory, is crucial. STREAM algorithm based on Divide and Conquer that achieves a constant factor approximation in small space. This STREAM algorithm is based on a facility location algorithm that might produce more than k centers. The advantage of STREAM algorithm is trade off between cluster quality and running time. This algorithm is compared with BIRCH Algorithm and proved that BIRCH appears to do a reasonable quick and dirty job.

Ashish Singhal and Dale E. Seborg together in [9] calculated the degree of similarity between multivariate time series datasets using two similarity factors with batch fermentation algorithm. One similarity factor is based on principal component analysis and the angles between the principal component subspaces. Second similarity factor is belongs to Mahalanobis distance between the datasets. Batch fermentation algorithms are to compare the product quality data for different datasets. The advantage of this similarity factor with batch fermentation is very effective in clustering multivariate time series datasets and is better to existing methodologies. It provides best clustering performance and the results are very close to each other and also the clustering performance is sensitive.

Hui Zhang et al. in [11] put forth an unsupervised feature extraction algorithm using orthogonal wavelet transform for automatically choosing the dimensionality of features. The problem of determining the feature dimensionality is circumvented by selecting the appropriate scale of the wavelet transform. When the dimensionality is reduced the information may be lost. This feature extraction algorithm controls the lower dimensionality and lower errors by choosing the scale within which the nearest lower scale. The major advantage of this feature extraction is chosen automatically. And the qualities of clustering with extracted features are better than that with features corresponding to the scale prior and posterior scale averagely for the used data sets.

Bagnall et al. in [10] explained a technique in order to assess the effects of dimensionality data into binary sequences of above and below the median, this process is known as clipping. For long time series data the clustering accuracy when using clipped data from the class of ACMA models is not significantly different to that achieved with unclipped data. The usage of clipped data produces better clusters, whether the data contains outliers, when using clipped data needs less memory and operations. And distance calculations can be much faster. Calculating auto corrections are faster with clipped data. Clipped data with clustering provides good clustering. But the data sets are massive automatically the execution speed of clustering algorithm is reduced.

Ernst et al. in [12] described an algorithm for clustering short time series gene expression data. Most clustering algorithms are not capable to make a distinction between real and random patterns. They presented an algorithm specifically designed for clustering short time series expression data. Their algorithm works by assigning genes to a predefined set of model profiles that capture the potential distinct patterns that can be expected from the experiment. They also discussed how to obtain such a set of profiles and how to determine the significance of each of these profiles. Significant profiles are retained for further analysis and can be combined to form clusters. Also they tested their method on both simulated and real biological data. Using immune response data they showed that their algorithm can correctly detect the temporal profile of relevant functional categories. Using Gene Ontology analysis the results showed that their algorithm outperforms both general clustering algorithms and algorithms designed specifically for clustering time series gene expression data.

A new clustering method for time series data streams was proposed by Li et al. in [14]. Clustering streaming time series is a complicated crisis. The majority of the traditional algorithms are too disorganized for large amounts of data and outliers in them. In their paper, they proposed a new clustering method, which clusters Bi-clipped (CBC) stream data. It contains three phrases, namely, dimensionality reduction through piecewise aggregate approximation (PAA), Bi-clipped process that clipped the real valued series through bisecting the value field, and clustering. Through related experiments, they found that CBC gains higher quality solutions in less time compared with M-clipped method that clipped the real value series through the mean of them, and unclipped methods. This situation is especially distinct when streaming time series contain outliers.

A clustering algorithm for time series data was put forth by Jian et al. in [15]. In the Intelligent Traffic System, the research about the analysis of time series of traffic flow is significant and meaningful. Using clustering methods to investigate time series not only can find some typical patterns of traffic flow, but also can group the sections of highway by their different flow characteristics. In their paper, they proposed an Encoded-Bitmap-approach-based swap method to improve the classic hierarchical method. Moreover, their experimental results showed that their proposed method has a better performance on the change trend of time series than classic algorithm.

Beringer et al. in [16] put forth a clustering algorithm for parallel data streams. In modern years, the management and processing of so-called data streams has become a subject of dynamic research in numerous fields of computer science such as, e.g., distributed systems, database systems, and data mining. A data stream can approximately be thought of as a transient, continuously increasing sequence of time-stamped data. In their paper, they considered the problem of clustering parallel streams of real-valued data, that is to say, continuously evolving time series. In other words, they are interested in grouping data streams the evolution over time of which is comparable in a specific sense. In order to maintain an up-to-date clustering structure, it is indispensable to investigate the incoming data in an online manner, tolerating not more than a constant time delay. For this purpose, they developed a resourceful online version of the classical K-means clustering algorithm. Their method's efficiency is mainly due to a scalable online transformation of the original data which allows for a fast computation of approximate distances between streams.

Characteristics based clustering of time series data was described by Wang et al. in [17]. Their paper proposed a method for clustering of time series based on their structural characteristics. Unlike other alternatives, their proposed method does not cluster point values using a distance metric, rather it clusters based on global features extracted from the time series. The feature measures are obtained from each individual series and can be fed into random clustering algorithms, including an unsupervised neural network algorithm, self-organizing map, or hierarchal clustering algorithm. Global measures describing the time series are obtained by applying statistical operations that best capture the underlying uniqueness: trend, seasonality, periodicity, serial correlation, skewness, kurtosis, chaos, nonlinearity, and self-similarity. Since the method clusters using extracted global measures, it reduces the dimensionality of the time series and is much less sensitive to missing or noisy data. They further provide a search mechanism to find the best selection from the feature set that should be used as the clustering inputs. Their technique has been tested using benchmark time series datasets formerly reported for time series clustering and a set of time series datasets with known distinctiveness. The empirical results show that their approach is able to yield meaningful clusters. The resulting clusters are comparable to those produced by other methods, but with some promising and interesting variations that can be instinctively explained with knowledge of the global characteristics of the time series.

Hirano et al. in [18] proposed an algorithm for clustering the time series medical data. Their paper presents a cluster analysis method for multidimensional time-series data on clinical laboratory examinations. Their method represents the time series of test results as trajectories in multidimensional space, and compares their structural similarity by using the multiscale comparison technique. It enables us to find the part-to-part correspondences between two trajectories, taking into account the relationships between different tests. The resultant distinction can be further used with clustering algorithms for finding the groups of similar cases. The method was applied to the cluster analysis of Albumin-Platelet data in the chronic hepatitis dataset. The experimental results demonstrated that it could form interesting groups of cases that have high correspondence to the fibrotic stages.

Clustering of time series clipped data was projected by Bagnall et al. in [19]. They showed that the simple procedure of clipping the time series reduces memory requirements and considerably speeds up clustering without decreasing clustering accuracy. They also demonstrated that clipping increases clustering accuracy when there are outliers in the data, thus serving as a means of outlier detection and a method of identifying model misspecification. They considered simulated data from polynomial, autoregressive moving average and hidden

Markov models and showed that the estimated parameters of the clipped data used in clustering tend, asymptotically, to those of the unclipped data. Moreover they demonstrated experimentally that, if the series are long enough, the accuracy on clipped data is not significantly less than the accuracy on unclipped data, and if the series contain outliers then clipping results in significantly better clusterings. Finally, they illustrated how using clipped series can be of practical benefit in detecting model misspecification and outliers on two real world data sets: an electricity generation bid data set and an ECG data set.

Nakamoto et al. in [20] explained a fast clustering algorithm for time series data. Their paper proposed a fast clustering method for time-series data based on a data structure: TWS (Time Warping Squashing) tree. A clustering procedure based on an comprehensive search method is time-consuming although its result exhibits high quality. BIRCH, which reduces the number of examples by data squashing based on a data structure: CF (Clustering Feature) tree, represents an efficient solution for such a method when the data set consists of numerical attributes only. For time-series data, on the other hand, a straightforward application of BIRCH based on a Euclidean distance for a pair of sequences, despondently fails since such a distance typically differs from human's perception. A distance based on DTW (Dynamic Time Warping) is advantageous, but no methods have been proposed for time-series data in the context of data-squashing clustering. In order to get around this problem, they proposed the TWS tree, which employs a distance based on DTW, and compresses sequences to an average sequence. An average sequence is obtained by a novel procedure which estimates correct shrinkage of a result of DTW. Experiments based on the Australian sign language data demonstrated the superiority of the proposed method in terms of correctness of clustering, while its deprivation of time-efficiency is insignificant.

Clustering of time series data gene expression data using smoothing spline derivatives was developed by Dejean et al. in [21]. Microarray data obtained during time-course experiments permit the temporal variations in gene expression to be monitored. An imaginative postprandial fasting experiment was conducted in the mouse and the expression of 200 genes was monitored with a dedicated macroarray at 11 time points between 0 and 72 hours of fasting. The intention of their study was to provide a applicable clustering of gene expression temporal profiles. This was achieved by focusing on the shapes of the curves rather than on the supreme level of expression. In point of fact, they combined spline smoothing and first derivative computation with hierarchical and partitioning clustering. A heuristic approach was proposed to tune the spline smoothing parameter using both statistical and biological considerations. Clusters are illustrated a posteriori through principal component analysis and heatmap visualization. Most results were found to be in agreement with the literature on the effects of fasting on the mouse liver and provide promising directions for future biological investigations.

## III. FUTURE WORK

Clustering time series data is a difficult task in the applications that has wide-range assortment of fields, and has recently attracted a large amount of research. The proposed study provides a way to investigate the existing algorithms and techniques for clustering of time series data streams and helps to give directions for future enhancement. Future research can be directed to the following aspects:

1. Cluster time series data in high dimensional data by increasing the speed.
2. Computation effort can be increased in high – dimensional data using clipping technique.
3. An effective approach can be developing to predict the future value in time series data.
4. Since, Time series data deals with raw format which is expensive in terms of processing and storage. In the proposed work a proposed time series data format can be taken to solve the above problem.

## IV. CONCLUSION

In modern years, the management and processing of so-called data streams has become a subject of dynamic research in numerous fields of computer science such as, e.g., distributed systems, database systems, and data mining. Lot of research work has been carried in this field to develop an efficient clustering algorithm for time series data streams. Time series data are frequently large and may contain outliers. Therefore, careful examination of the earlier proposed algorithms is necessary. In this paper we surveyed the current studies on time series clustering. These studies are structured into many categories depending upon whether they work directly with the innovative data. Most clustering algorithms are not capable to make a distinction between real and random patterns. In addition, this paper discusses about possible high dimensional problems with time series data. The application areas are summarized with a brief description of the data used. The uniqueness and drawbacks of past studies and some possible topics for further study are also discussed. The future work determines to develop an effective clustering algorithm for time series data streams.

| | |
|---|---|
| 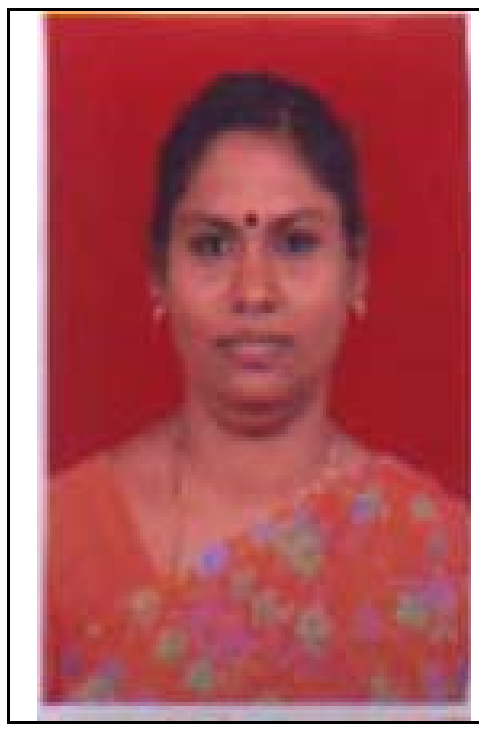 | **Dr. M.Punithavalli** is presently working as Director & Head of the Dept of Computer Science, Sri Ramakrishna College of arts and science for women College, India.She has published more than twenty papers in national/International journals. Her areas of interest includes E-Learning, Software Engineering, Data Mining, Networking and etc. She has about 16 years of teaching experience. She is guiding many research scholars and has published many papers in national and international conference and in many international journals |
| 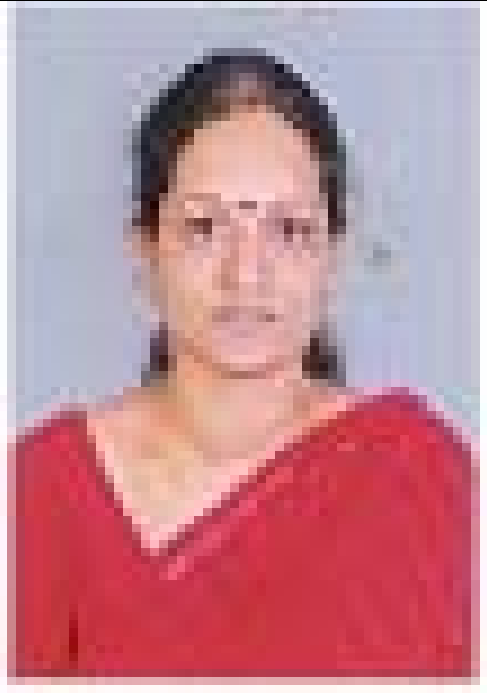 | **V.Kavitha** graduated with M.C.A in 2001 from Bharathidasan University, India and completed M.Phil from Alagappa University, India during 2005. Her area of Interest include Data Mining. She has about 7 years of teaching experience. Currently she is working as a Lecturer in Sri Ramakrishna college of Arts & Science for Women,India. She has presented a paper in two national conferences . |